\documentstyle[emulateapj,apjfonts,psfig]{article}

\lefthead{Miller \& Homan}  
\righthead{The Iron Line -- QPO Connection}

\received{Date}
\revised{Date}
\accepted{Date}

\journalid{vol}{date}
\articleid{1}{4}
\paperid{id}

\cpright{AAS}{1999}
\ccc{x}

\begin{document}

\title{Evidence for a Link Between Fe K$\alpha$ Emission Line Strength and
  QPO Phase in a Black Hole}

\author{J.~M.~Miller\altaffilmark{1,2} and J.~Homan\altaffilmark{3}}

\altaffiltext{1}{Harvard-Smithsonian Center for Astrophysics, 60
	Garden Street, Cambridge, MA 02138, jmmiller@cfa.harvard.edu}
\altaffiltext{2}{NSF Astronomy and Astrophysics Fellow}
\altaffiltext{3}{Center~for~Space~Research and Department~of~Physics,
        Massachusetts~Institute~of~Technology, Cambridge, MA
        02139--4307, jeroen@space.mit.edu}

       
\authoremail{jmmiller@cfa.harvard.edu}

\label{firstpage}

\begin{abstract}
In X-ray binaries, the frequencies revealed in X-ray quasi-periodic
oscillations (QPOs) are often interpreted as characteristic
frequencies in the inner accretion disk, though the exact oscillation
mechanism is unknown at present.  Broadened Fe~K$\alpha$ lines are
also excellent diagnostics of the inner accretion flow, if their
broadening is indeed due to inner disk reflection.  Herein, we present
two cases where the flux and equivalent width of the Fe~K$\alpha$
emission lines in spectra of the Galactic black hole GRS~1915$+$105
vary with the phase of strong 1~Hz and 2~Hz QPOs in the X-ray flux.
These results provide strong evidence that both QPOs and the
Fe~K$\alpha$ lines originate in the inner disk.  If the 1~Hz QPO is
only a Keplerian orbital frequency, the QPO comes from a distance of
$84\pm 26~R_{Schw.}$ from the black hole; the 2~Hz QPO implies a
radius of $50\pm 15~R_{Schw}$.  At these radii, relativistic shaping
of a disk line is inevitable.  Moreover, the link holds in
radio--bright and radio--faint phases, signaling that in systems like
GRS 1915$+$105, the Fe~K$\alpha$ line is a disk line and not a jet
line as per SS~433.  A particularly interesting possibility is that a
stable warp in the inner disk, e.g. due to Lense-Thirring precession,
may produce the observed QPOs and line modulations. More broadly, the
Fe~K--QPO link provides an unprecedented mechanism for revealing the
inner accretion flow and relativistic regime in accreting systems, in
that it gives two measures of radius: for a given disk QPO model, the
frequency translates into a specific radius, and relativistic line
models yield radii directly.
\end{abstract}

\section{Introduction}
In accretion-powered sources, a great deal of theoretical and
observational effort is devoted to the study of the inner accretion
flow, and to using the accretion flow itself as a tool to study the
central compact object.  High frequency QPOs in Galactic black hole
systems may be related to the Keplerian frequency at the innermost
stable circular orbit (for zero-spin black holes, $\nu = 220~Hz~
10~M_{\odot}/M_{BH}$; e.g., Strohmayer 2001, Miller et al.\ 2001,
Remillard et al.\ 2002, Homan et al.\ 2004), but they are only
observed in a handfull of sources.  QPOs at lower frequencies,
typically 1--10 Hz are much more common.  Although their origin is
unclear, interpreting them as a Keplerian frequency puts them at
$\lesssim$100 $R_{Schw.}$. If these QPOs are a precession frequency at
the inner disk (e.g., due to frame dragging, see Markovic \& Lamb
1998), they probe regions much closer to the black hole. Broad,
asymmetric Fe~K$\alpha$ emission lines are found in the spectra of
both supermassive black holes and stellar-mass Galactic black holes
(Reynolds \& Nowak 2003).  These lines appear to be imprinted with the
extreme Doppler shifts and gravitational red-shifts expected near to a
black hole, but the case is not entirely shut.

A natural evolution of the means by which inner accretion flows are
studied is to bring spectral and timing constraints to bear jointly.
It has been shown that the disk reflection spectrum (the Fe~K$\alpha$
line is the most prominent part of the reflected spectrum; see George
\& Fabian 1991) in Galactic black holes varies with broad-band
frequency (Gilfanov, Churazov, \& Revnivtsev 2000).  This is strong
evidence that timing and spectral properties are intimately related,
and supports models wherein Fe~K$\alpha$ emission lines arise in the
inner disk.

In this work, we take spectral and timing connections a step further.
We report the discovery of a link between discrete timing features
(QPOs) and spectral features (Fe~K$\alpha$ emission lines):
Fe~K$\alpha$ emission line strength varies with QPO phase in
radio--bright and radio--faint states of the ``microquasar''
GRS~1915$+$105.  Our results provide the first indication that these
diagnostics of the inner accretion flow are related.

\section{Data Selection and Reduction}
We chose to investigate GRS~1915$+$105 as strong QPOs (see, e.g.,
Morgan, Remillard, \& Greiner 1997) and Fe~K$\alpha$ emission lines
(Muno et al.\ 2001, Martocchia et al.\ 2002, Miller et al.\ 2004a)
have been reported in this source.  The high mass of GRS 1915$+$105
($M_{BH} = 14.4\pm 4.4 M_{\odot}$, Harlaftis \& Greiner 2004) means
that for a given QPO frequency, the QPOs may trace regions closer to
the black hole in this source than in lower-mass black holes.
GRS~1915$+$105 displays a wide variety of radio states.  Steady radio
emission is associated with compact jets (Fender 2004), and this fact
can be exploited to examine whether any relation between Fe~K$\alpha$
line flux and QPOs depends on the presence of a jet.

To extract QPO phase-selected spectra, QPOs should be strong enough to
dominate the X-ray variability and be directly visible in the X-ray
lightcurves.  Morgan, Remillard, \& Greiner (1997) found
large amplitude, sinusoidal QPOs in GRS~1915$+$105 that could be seen
in the naked lightcurve.  The QPOs were found to be quasi--periodic
largely due to random jumps in phase.  We selected two observations of
GRS~1915$+$105 available in the {\it RXTE} public archive:
20402-01-15-00 (obs.\ date: 9 Feb.\ 1997) and 20402-01-50-01
(obs. date: 16 Oct.\ 1997).  The spectral and timing properties of
these observations (including an Fe~K$\alpha$ emission line)

\centerline{~\psfig{file=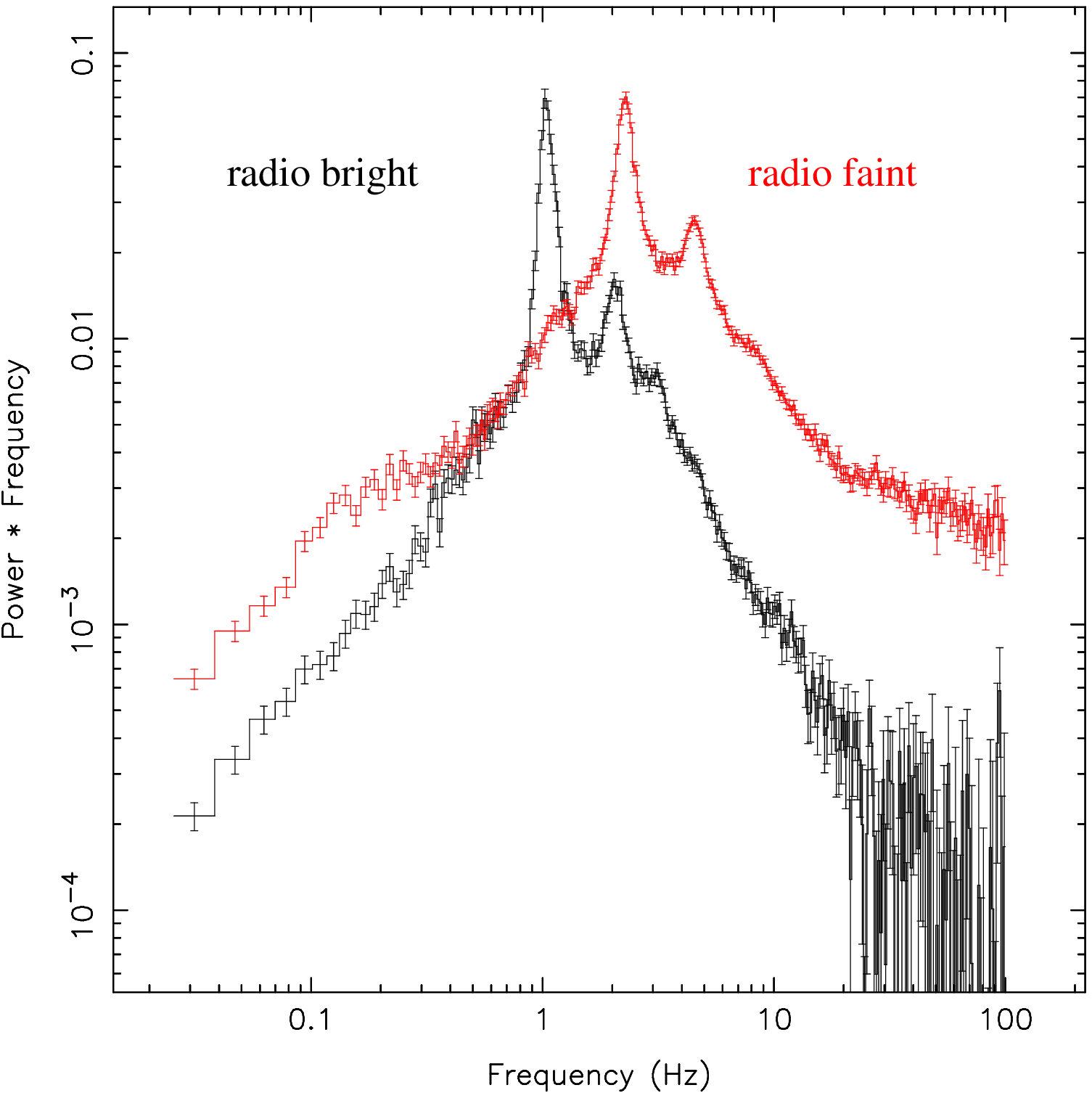,width=3.0in}~}
\figcaption[h]{\footnotesize The power density spectra of the
  radio-bright observation (black) and radio-faint observation (red)
  are shown above.
  The strongest QPOs have frequencies of 1.052(3)~Hz and 2.260(2)~Hz,
  and amplitudes of 14.7(3)\% rms and 13.6(1)\% rms, respectively.}
\smallskip

\noindent have previously been reported on by Muno et al.\ (2001).
Importantly, Muno et al.\ (2001) also studied the radio properties of
GRS~1915$+$105 during these observations; the first occurred in a
``radio--faint'' state (hereafter, ``RF''; the source was not detected
at 8.3 GHz), and the second in a radio-bright or radio-plateau state
($74\pm 6$ mJy at 8.3~GHz; hereafter ``RB'').

Tools within LHEASOFT version 5.3 and CIAO version 3.1 were used to
reduce and analyze these observations.  Standard {\it RXTE} data
screening and good time selections were applied (e.g., for SAA
avoidance).  We further screened the data to only include times during
which all five PCUs were operated.  The net total good time for the RF
obs.\ (15-00) was 10.1~ksec, and the net total good time for RB
obs.\ (50-01) was 4.5 ksec.  All spectra and lightcurves were
generated using the tool ``saextrct'', and all response matrices were
generated with the tool ``pcarsp''.  As GRS~1915$+$105 is very bright
and we are interested in relative variations, background spectra were
not subtracted.  We made time-averaged broad-band PCA spectra by
combining data from all layers of all PCUs taken in ``Standard-2''
mode (129 energy channels between 2 and 60 keV, taken every 16s).
Spectra and lightcurves were also made from data taken in
``B\_8ms\_16A\_0-35\_H'' mode (16 energy channels up to channel 35 --
roughly 13 keV -- taken every 8ms).  

To accurately estimate {\it relative} flux differences in data
acquired with the same detector, it is only important to understand
the degree to which systematic errors may {\it change} over time.  For
this purpose, we reduced and analyzed Crab spectra obtained before and
after each observation of GRS~1915$+$105 (spectra obtained on 31 Jan.\
1997 and 16 Feb.\ 1997, and 12 Oct.\ 1997 and 27 Oct.\ 1997).
Absorbed power-law fits to these spectra reveal that most channels
below 10~keV (those most important for Fe K$\alpha$ line studies)
differ by 0.2\% or less.  To be conservative, we added 0.2\%
systematic errors to the spectra.  This is less than the 0.5--1.0\%
systematic errors often added to PCA spectra when absolute fluxes are
of interest.

\section{Analysis and Results}
We made power-density spectra (PDS) of the X-ray flux from each
observation using all available data, and fit the 

\centerline{~\psfig{file=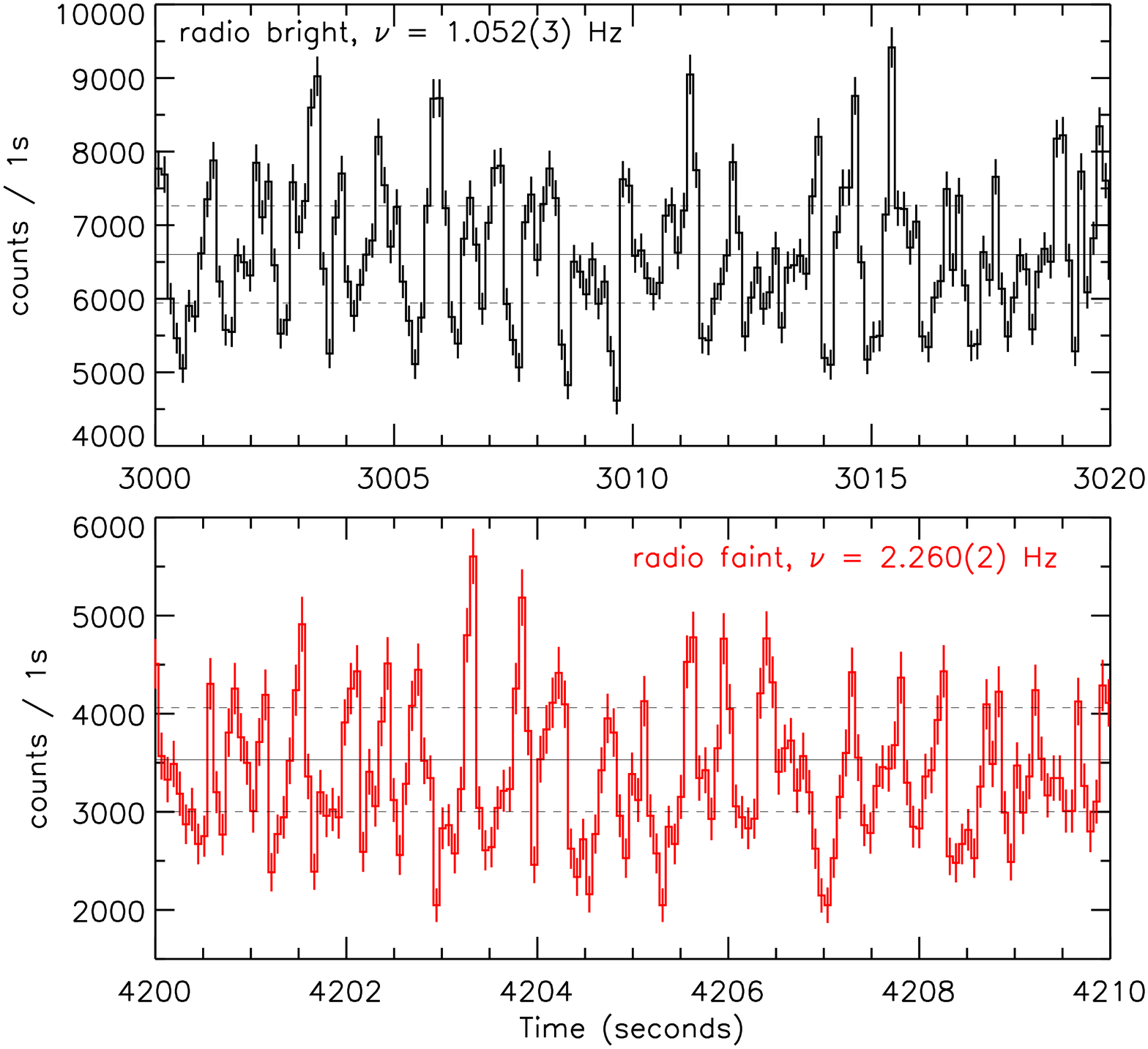,width=3.2in}~}
\figcaption[h]{\footnotesize Approximately 20 cycles of the QPO wave
  from randomly selected intervals in each observation are shown
  above.  Clearly the QPOs strongly dominate the lightcurves.  Solid
  lines mark the mean count rate, and dashed lines above and below
  mark the count rate levels used to select the high and low phases of
the QPO waves.  The radio-bright data are binned to 0.128~s, and the
radio-faint data are binned to 0.064~s.}
\smallskip

\noindent 0.01--100~Hz range with Lorentzian functions in the
$\nu$--max representation (see, e.g., Homan et al.\ 2004).  Each PDS
is strongly dominated by a single QPO (see Figure 1).  The strongest
QPO in the RF obs.\ has a frequency of 2.273(4) Hz, a $Q$--value of
6.0(2), and an amplitude of 11.8(2)\% rms (errors are 1$\sigma$).  The
strongest QPO in the RB obs.\ has a frequency of 1.050(3) Hz, a
$Q$--value of 7.9(4), and an amplitude of 11.1(2)\% rms.  Assuming a
black hole mass of $M_{BH} = 14.4\pm 4.4~M_{\odot}$, if these
frequencies are merely Keplerian orbital frequencies, they correspond
to radii of $84\pm 26~R_{Schw.}$ and $50\pm 15~R_{Schw.}$
respectively.

Lightcurves from the ``B\_8ms\_16A\_0-35\_H'' mode data from each
observation were analyzed, and the total mean count rate and mean
count rate in 100s intervals were calculated.  The mean rate in
the RF obs.\ is 3532 c/s, with a range of 3298--3727 c/s when measured
in 100s intervals.  The mean rate in the RB obs.\ is 6602 c/s,
with a range of 6496--6793 c/s.

As the QPOs in each observation dominate the X-ray variability (see
Figure 2), and because the mean count rate in each observation is very
steady, we applied simple count-rate selections to isolate the maxima
and minima of the QPO waves.  For the RF obs.\, selecting periods when
the flux was 15\% above or below the mean effectively isolated the
maxima and minima.  For the RB obs.\, there is less noise apart from
the QPO, and periods when the flux was 10\% above or below the mean
were found to effectively isolate maxima and minima.  The CIAO tool
``dmgti'' was used to generate additional good time files of the
maxima and minima.  These files were than applied within ``saextrct''
to produce spectra of the maxima and minima.

All spectral analysis was done using XSPEC version 11.3 (Arnaud 1996)
and IDL version 5.4.  Analysis of the time--averaged standard-2
spectra was performed in the 3.0--20.0 keV band.  The PCA is
calibrated well in this energy range, and this range is not much
greater than the range of the ``B\_8ms\_16A\_0-35\_H'' mode data. Fits
to the spectra with a number of common models revealed the
Fe~K$\alpha$ emission line previously noted by Muno et al.\ (2001).
In each case, the addition of a simple 

\centerline{~\psfig{file=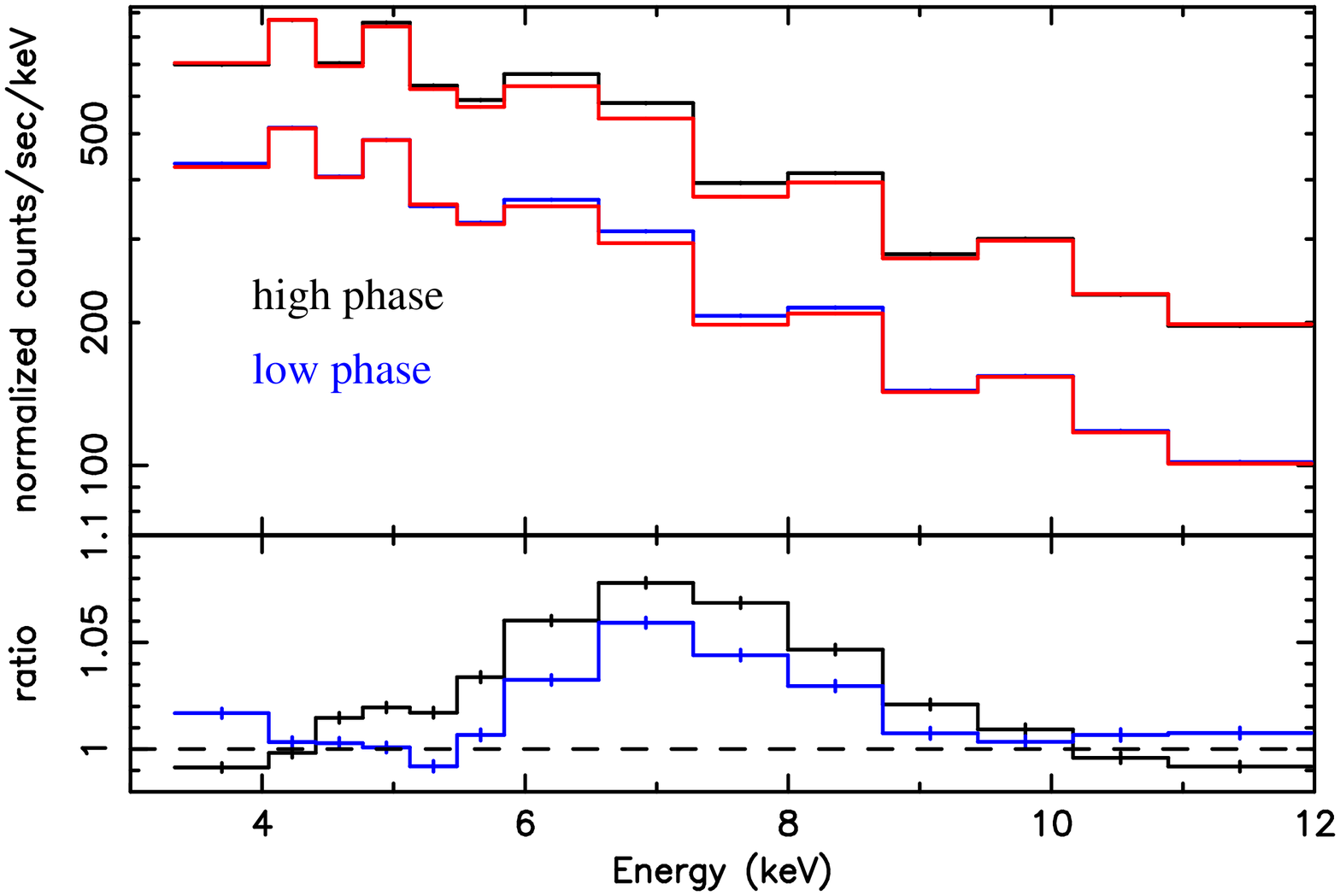,width=3.2in}~}
\figcaption[h]{\footnotesize Fits to the high and low phase spectra
  from the radio--faint observation of GRS~1915$+$105 are shown
  above.  The spectra were fit with the model described in Table 1,
  using the fixed line centroid and width (which gives a smaller
  difference in the line fluxes than if these parameters float).  The
  line flux was set to zero on the ratio plot to show the difference
  in Fe~K$\alpha$ line intensity between high and low QPO phases.}
\smallskip

\noindent Gaussian component to model the line was significant at more
than the 8$\sigma$ level of confidence.  We find that disk reflection
models provide the best overall fit to the broad-band spectrum and do
not require a disk that is prominent in the {\it RXTE} bandpass.  Our
model does not include a disk, but this only reflects the fact that a
cool disk with moderate flux is not easily seen in the {\it RXTE}
bandpass, especially when $N_{H}$ is high.

We fit the time-averaged, high phase, and low phase spectra with
absorbed cut-off power-law (for the radio-bright phase) or broken
power-law (for the radio-faint phase) models, modified by the addition
of a Gaussian component (to model the Fe~K$\alpha$ emission line) and
smeared edge component (``smedge'' in XSPEC) to mimic a reflected
continuum.  Trudolyubov (2001) also found that these differing
spectral models were required in RF and RB states.  The ``phabs''
model was used to account for absorption in the neutral ISM.  The
``smedge'' was not statistically required in all fits, and its
parameters could not be well constrained.  However, the smedge
effectively mimics a disk reflection spectrum at energies above the
Fe~K$\alpha$ emission line, so we fixed the model parameters to modest
values consistent with the data ($E = 8.5$~keV, $\tau = 0.3$,
width$=10$~keV).

The parameters obtained from the spectral fits are shown in Table 1.
In all cases, the line fluxes and equivalent widths are significantly
different in the high and low phase spectra (see Fig.\ 3).  The
confidence intervals do not overlap even when the 1$\sigma$ error bars
are multiplied by 5.  There is evidence that the Fe~K$\alpha$ emission
line flux is broader and slightly shifted to the red in the high phase
relative to the low phase.

To demonstrate that the line flux varies in a model--independent
manner, we subtracted the low count rate spectra from the high count
rate spectra, and divided the result by the mean count rate for the
entirety of each observation.  Using the energy to channel bounds found in
the response matrix for each observation, the bins in the difference
spectra were converted to energy bounds; however, the detector
response was not removed and no spectral fitting was performed.  In
both the radio--bright and radio--faint observations, the most
prominent difference between the high and low phases is the
Fe~K$\alpha$ line (see Fig.\ 4).  For a ``control'' observation, we
extracted high and low phase spectra (10\% above/below the mean) from
a 1.2~ksec slice of an observation of the Galactic black hole
H~1743$-$322 obtained on 28 May 2003.  Like the observations of
GRS~1915$+$105, H~1743$-$322 was highly variable on 

\centerline{~\psfig{file=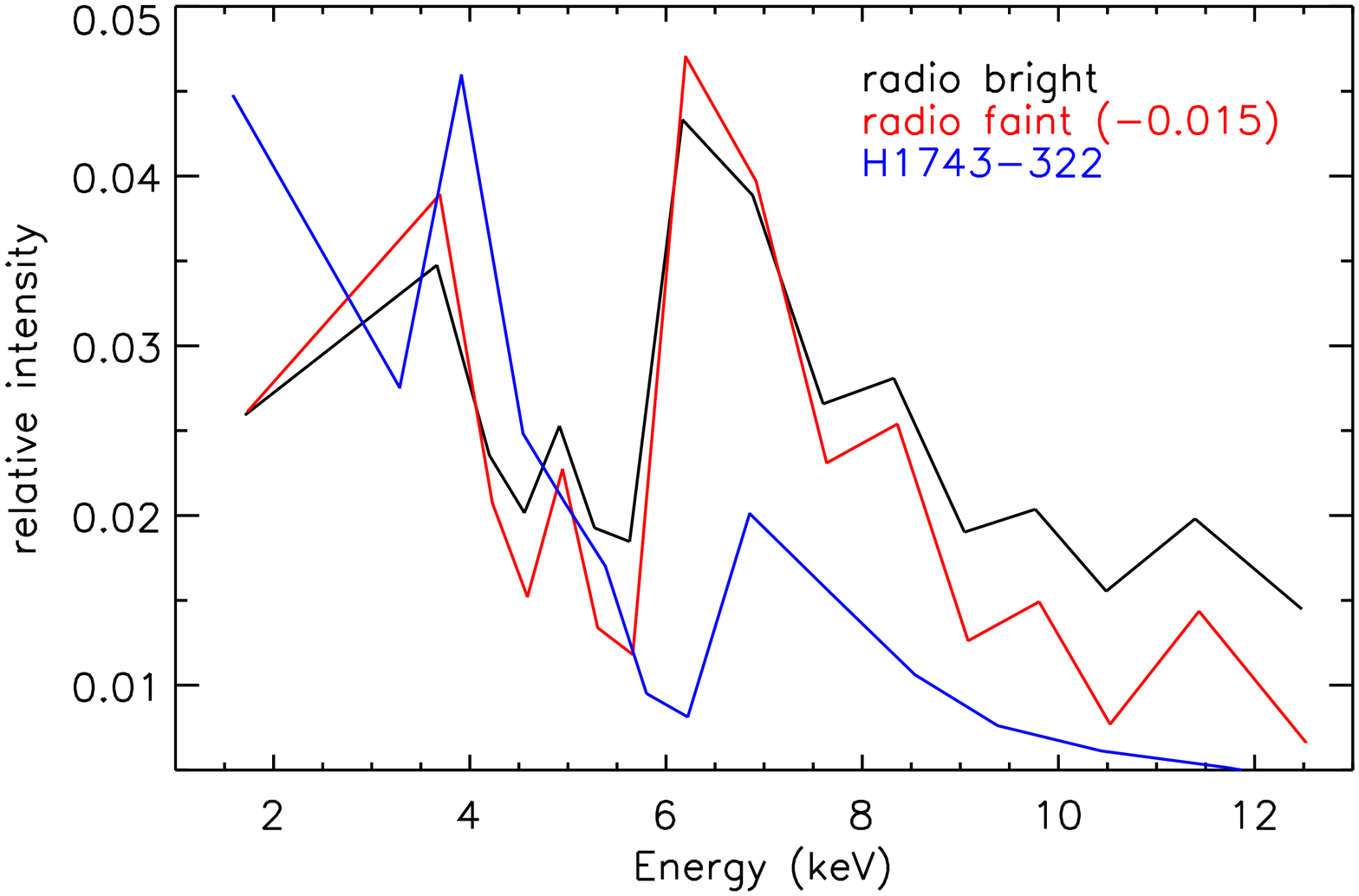,width=3.2in}~}
\figcaption[h]{\footnotesize The plot above shows the relative
  intensity of the radio-bright and radio-faint spectra of
  GRS~1915$+$105 (and a ``control'' observation of H~1743$-$322) in
  high and low count rate phases.  The low count rate spectra were
  subtracted from the high count rate spectra, and the difference was
  divided by the mean count rate for the entire observation.  Error
  bars are plotted but are very small.  Channel bins were converted to
  energy space but the detector response has not been removed (so the
  Xe L edge complex between 4--6 keV is seen).  Unlike the
  H~1743$-$322 difference spectrum, the Fe~K$\alpha$ line bins are
  dominant in the GRS 1915$+$105 difference spectra.  This
  model--independent analysis confirms that the Fe~K$\alpha$ line
  intensity varies with QPO phase.}
\smallskip

\noindent timescales of seconds, but had a steady mean count rate on
100~s scales in this window.  Spectroscopy of H~1743$-$322 in outburst
did not reveal an Fe~K$\alpha$ emission line (Miller et al.\ 2004b),
and the Fe~K$\alpha$ line bins are not globally dominant in the
difference spectrum shown in Fig.\ 4.

\section{Discussion and Conclusions}
The main results of this work may be summarized as follows:\\
\noindent $\bullet$ Fe~K$\alpha$ line flux (and possibly FWHM and centroid
energy) depends on QPO phase in 1--2~Hz QPOs in GRS~1915$+$105,
linking two independent diagnostics of the inner accretion flow.\\
$\bullet$ The QPO frequencies at which the Fe~K$\alpha$ line flux is
modulated correspond to radii of $84\pm 26~R_{Schw.}$ and $50\pm
15~R_{Schw.}$, if they are merely Keplerian orbital frequencies.
This link likely ties Fe~K$\alpha$ lines in Galactic black holes to
the inner disk and further supports evidence that the observed line
broadening is due to dynamics in the disk (Doppler shifts,
gravitational red-shifts) close to the black hole.\\
$\bullet$ The presence of Fe~K$\alpha$ lines, QPOs, and the link
between them does not depend on jet activity (taking radio flux as a jet
indicator).  Although the base of a failed jet may be a source of
hard X-rays which irradiate the disk, Fe~K$\alpha$ lines and QPOs
do not arise in an extended jet.\\

Blobs in the inner accretion disk might create QPOs and could give a
quasi-periodically changing reflector area.  Similarly, if the height
of hard X-ray emitters (flares, or the base of a jet) changes
quasi-periodically, QPOs and Fe~K$\alpha$ line variations might be
expected.  However, while both explanations are possible, such
explanations seem inconsistent with the strength and constancy of the
QPOs found in these observations of GRS~1915$+$105, and with the fact
that the Fe~K--QPO connection holds over a factor of two in flux and
in both radio-bright and radio-faint phases.  Moreover, associating
low-frequency QPOs with Keplerian frequencies is problematic for two
reasons: (1) X-ray production in accretion disks should be centrally
concentrated, with little emission from $r \geq 100~R_{Schw.}$, and
(2) QPOs are stronger at higher energies, suggesting they are also
centrally produced (see, e.g., Homan et al.\ 2001).  A stable warp at
the inner disk, e.g. due to Lense-Thirring precession, could
cause QPOs and would also serve to create a quasi-periodically
changing reflector area, modulating the strength of Fe~K$\alpha$ line
emission.  It has been suggested that Lense-Thirring precession may be
the cause of low-frequency QPOs in accreting systems (e.g., Markovic
\& Lamb 1998), and the Fe~K--QPO connection may be evidence
for this General Relativistic effect.

Whether or not Lense-Thirring precession can account for the Fe~K--QPO
connection, the Fe~K--QPO connection has the potential to reveal the
innermost regime in accreting systems.  For a given disk--driven QPO
model, the Fe~K--QPO connection gives two measures of radius:
frequencies imply specific radii, and relativistic disk line models
constrain radii directly (Fabian et al.\ 1989, Laor 1991).  The
connection over--constrains the system, if QPOs are disk frequencies.

Work of this kind in Galactic black holes may have important
implications for Fe~K$\alpha$ line reverberation mapping in AGN with
planned missions such as {\it Constellation-X} and {\it XEUS} (e.g.,
Young \& Reynolds 2000), and observations of disk hot spots orbiting
close to the black holes in AGN (e.g., Iwasawa, Miniutti, \& Fabian
2004).  Long {\it RXTE} exposures with carefully chosen PCA modes are
needed to obtain additional data for new studies.  The large
effective areas of {\it XMM-Newton} and {\it Astro-E2} may support
such studies as well, and may even allow studies of the line shape
with QPO phase.  A new X-ray mission with an area of 2--10m$^{2}$ and
CCD spectral resolution (or better) may fully exploit the Fe~K--QPO
connection.\\

We wish to thank Ed Bertschinger, Andy Fabian, Keith Jahoda, Fred
Lamb, Cole Miller, Ed Morgan, and Mike Nowak for helpful discussions.
We thank Keith Arnaud, Aneta Siemiginowska, and Tod Strohmayer for
guidance in handling systematic errors in this work.
J.M.M. acknowledges funding from the NSF, through its AAPF program.
This work was partially supported by funding from the {\it RXTE} GO
program.

\begin{table}[t]
\caption{Fits to QPO Phase Spectra}
\begin{footnotesize}
\begin{center}
\begin{tabular}{llllllllll}
~   & $N_{H}$ & $\Gamma_{1},\Gamma_{2}$ & E$_{break}$ (keV) & Norm. & E$_{Gauss}$ (keV)
  & FWHM & flux ($10^{-2}$ ph/cm$^{2}$/s) & EW (eV) & $\chi^{2}/\nu$\\
\tableline
RF avg. & 5.40(6) & 2.18(1),1.922(5) & 11.4(1) & 4.55(6) & 6.58(5) & 1.9(1) & 1.13(7) & 150(10) & 70.5/38.0 \\
RF high$^{\dag}$ & 5.40 & 2.125(3),2.5(1) & 11.4 & 5.45(3) & 6.58 & 1.9 & 2.00(8) & 200(8) & -- \\
RF low$^{\dag}$ & 5.40 & 2.335(6),1.00(5) & 11.4 & 4.42(2) & 6.58 & 1.9 & 0.66(3) & 120(5) & -- \\
RF high & 5.40 & 2.122(4),2.3(1) & 11.4 & 5.37(4) & 6.32(6) & 2.7(2) & 2.8(2) & 260(20) & -- \\
RF low & 5.40 & 2.328(6),1.29(9) & 11.4 & 4.32(4) & 6.72(4) & 1.1(1) & 0.52(3) & 100(6) & -- \\
\tableline
~   & $N_{H}$ & $\Gamma_1$ & E$_{cut}$ (keV) & Norm. & E$_{Gauss}$ (keV)
  & FWHM & flux ($10^{-2}$ ph/cm$^{2}$/s) & EW (eV) & $\chi^{2}/\nu$\\
\tableline
RB avg. & 5.96(8) & 1.97(2) & 29(1) & 7.3(2) & 6.43(3) & 1.9(1) & 4.2(2) & 280(10) & 189.2/39 \\
RB high$^{\dag}$ & 5.96 & 1.927(2) & 29 & 8.27(3) & 6.43 & 1.9 & 5.6(1) & 300(5) & -- \\
RB low$^{\dag}$ & 5.96 & 2.044(1) & 29 & 6.89(2) & 6.43 & 1.9 & 3.13(4) & 250(10) & -- \\
RB high & 5.96 & 1.925(3) & 29 & 8.18(5) & 6.33(3) & 2.4(1) & 6.7(2) & 360(10) & -- \\
RB low & 5.96 & 2.046(1) & 29 & 6.96(2) & 6.53(1) & 1.4(1) & 2.59(4) & 220(4) & -- \\
\tableline
\end{tabular}
\end{center}
\tablecomments{``RF'' refers to the radio-faint observation (fit with
  a broken power-law), and ``RB'' to the radio-bright observation (fit
  with a cut-off power-law).  A ``smedge'' with E$=$8.5~keV,
  $\tau=0.3$, and width$=$10~keV was included in all fits.  Fits to
  ``avg'' spectra are fits to standard-2 data in the 3--20~keV band.
  All other fits are to lower resolution ``B\_8ms\_16A\_0-35\_H''
  data in the 3--13~keV band.  $^{\dag}$ Denotes fits in which the
  Fe~K$\alpha$ line centroid and FWHM were fixed to the values
  measured in the average spectrum.  Where errors are not given, parameters
  from the standard-2 fits were fixed.  Systematic errors of 0.2\%
  were added to all spectra prior to fitting to estimate the drift in
  detector response.  Fits to the phase-selected spectra with line
  centroid energy and FWHM fixed to the average value, and with
  parameters free, are both reported for comparison.  All errors are
  $1\sigma$ errors to allow confidences to be inferred directly.
  Formally acceptable fits are obtained with systematic errors set to
  0.6\%, which is typical for spectroscopy wherein absolute flux
  measurements are important.}
\vspace{-1.0\baselineskip}
\end{footnotesize}
\end{table}

\end{document}